# Controlled generation of array beams of higher order orbital angular momentum and study of their frequency doubling characteristics

B. S. Harshith[1,2,*] and G. K. Samanta[1]

[1]Photonic Sciences Laboratory, Physical Research Laboratory, Navrangpura, Ahmedabad, Gujarat 380009, India
[2] Indian Institute of Science Education and Research Pune, Dr. Homi Bhabha Road, Pashan, Pune, Maharashtra 411008, India
*Corresponding author: ssharshith.bachimanchi@gmail.com

**Abstract**: We report on a simple and compact experimental scheme to generate high power, ultrafast, higher order vortex array beams. Simply by using a dielectric microlens array (MLA) and a plano-convex lens we have generated array beams carrying the spatial property of the input beam. Considering the MLA as a 2D sinusoidal phase grating, we have numerically calculated the intensity pattern of the array beams in close agreement with the experimental results. Using Gaussian embedded vortex beams of order as high as $l = 6$, we have generated vortex array beams with individual vortices of order as high as $l = 6$. We have also theoretically derived the parameters controlling the intensity pattern, size and the pitch of the array and verified experimentally. The single-pass frequency-doubling of the vortex array at 1064 nm in a 1.2 mm long BiBO crystal produced green vortex array of order, $l_{sh} = 12$, twice the order of the pump beam. Using lenses of different focal lengths, we have observed the vortex array of all orders to follow the focusing dependent conversion similar to the Gaussian beam. The maximum power of the green vortex array is measured to be 138 mW at a single-pass efficiency as high as ~3.65%. This generic experimental scheme can be used to generate array beams of desired spatial intensity profile across wide wavelength range by simply changing the spatial profile of the input beam.

Optical vortices, doughnut shaped optical beams with helical wavefront, carry orbital angular momentum (OAM) per photon. Typically, the optical vortices are generated by impinging a helical phase factor $exp(il\varphi)$, where, $\varphi$ is the azimuthal angle and the $l$ is the topological charge or the order of a vortex beam, to the Gaussian beams with the help of spatial mode converters including spiral phase plates (SPPs) [1], q-plates [2], and holographic spatial light modulators (SLMs) [3]. Since their discovery, the vortex beams have found a great deal of attention for their wide range of applications in a variety of fields in science and technology including particle trapping and micro-manipulation [4], quantum information [5] and micromachining [6]. However, the recent advancements on multiple particle trapping and tweezing [7], fast micromachining [8], and multiplexing in quantum information [9] demand optical beams with vortex arrays in a simple experimental scheme. While majority of the existing mode converters transform the Gaussian beam into a single vortex beam, the intrinsic advantage of the dynamic phase modulation through holographic technique allow the SLMs to generate vortex arrays directly from a Gaussian beam [10]. However, the low damage threshold of SLMs restrict their usage for high power vortex array applications. As such one need to explore alternative techniques to generate array of vortex beams.

Efforts have been made to generate optical beamlet arrays using annular apertures with optical coherence lattices [11] and modified Fresnel zone plates [12]. In addition to the difficulty of manufacturing such devices, the power loss in such devices restrict their use for high power operations. On the other hand, plasmonic metasurface nano apertures [13] and amplitude gratings [14] have produced optical vortex arrays in Talbot planes through self-imaging. Unfortunately, the stringent dependence of the input grating parameters on the self-imaging, results in limited or no control on the properties of the vortex array. Here, we report a simple experimental scheme based on a dielectric microlens array (MLA) and a plano-convex lens to generate high power array beams carrying the spatial property of the input beam. Using ultrafast, high power vortex beams of orders as high as, $l = 6$, at 1064 nm in the input, we have generated vortex beam arrays and subsequently frequency-doubled into vortex arrays at a new wavelength. Using simple mathematical treatment, we have numerically calculated the intensity distribution of the vortex array and also derived the parameters controlling the vortex array in close agreement with the experimental results.

The schematic of the experimental setup is shown in Fig. 1. A 5 W Yb-fiber laser with spectral linewidth of 15 nm centered at 1064 nm providing femtosecond pulses of width ~260 fs at a repetition rate of 78 MHz is used as the pump laser. The input power to the setup is varied using a half waveplate (λ/2) and polarizing beam splitter cube (PBS1). The input beam is expanded and collimated using a telescopic combination of lenses, L1 and L2 of focal length, $f = 50$ mm and $f = 100$ mm, respectively. Using two spiral phase plates, SPP1 and SPP2, with phase winding corresponding to vortex orders $l = 1$ and $l = 2$, respectively and the vortex doubler [15] comprised with the PBS2, quarter-wave plate (λ/4) and mirror, M1, we have converted the Gaussian beam into vortex of order, $l = 1 - 6$. The pump vortex beam of order, $l_p$, on propagation through the lens L3 of focal length $f = 300$ mm, and the microlens array (MLA) (Thorlabs MLA 300-14AR) consisting with 391 lenslets of focal length, $f_{MLA} = 18.6$ mm resembling a 2D sinusoidal phase pattern, is Fourier transformed to produce vortex array at the back focal plane of the lens L3 [16]. The distance, $d$, of the MLA from the back focal plane of L3 modulates the pitch of the vortex array and the focal length, $f$ of the lens L3 determines the diameter of individual vortices in the array.

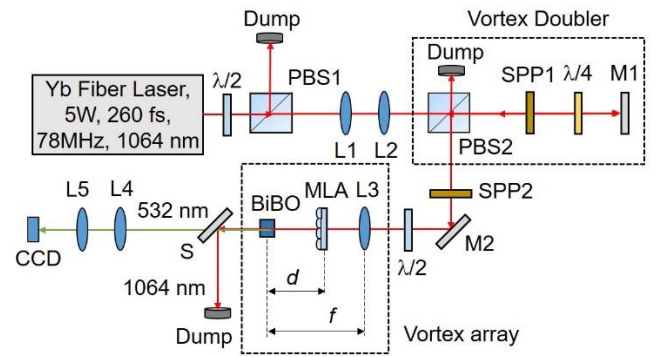

Fig. 1. Schematic of the experimental setup to generate vortex beam array. λ/2, half-wave plate; PBS1-2, polarizing beam splitter cube; SPP1-2, spiral phase plates; λ/4, quarter-wave plate; L1-5, lenses; M1-2, mirrors; BiBO, nonlinear crystal for frequency doubling; MLA, microlens array; S, wavelength separator; CCD, CCD camera.

A 1.2 mm long and 4 x 8 mm² in aperture bismuth borate (BiBO) crystal [17], cut for type-I (e + e → o) frequency doubling in optical yz-plane (Φ = 90°) with internal angle of θ = 168.5° at normal incidence, is placed at the Fourier plane for second harmonic generation (SHG) of vortex arrays at 1064 nm into green at 532 nm. The frequency doubled vortex array is extracted from the undepleted pump using a wavelength separator, S, and imaged at the CCD camera palne using lenses, L4 and L5.

To understand the formation of optical vortex arrays in the present experimental scheme, we have approximated the MLA as a 2D sinusoidal phase grating [18] of pitch, $\Lambda$, and sinusoid amplitude thickness, $s$ [19]. The transverse field amplitude distribution of the vortex beam of order, $l$, after the MLA can be written as [16]

$$E_{MLA} \sim exp\left[i\frac{m}{2}sin\left(\frac{2\pi x}{\Lambda}\right)\right] exp\left[i\frac{m}{2}sin\left(\frac{2\pi y}{\Lambda}\right)\right] E_l \quad (1)$$

where,

$$E_l \sim \left[\frac{(r\sqrt{2})^l}{(w_g)^{l+1}} exp\left(\frac{-r^2}{w_g^2}\right) exp(il\varphi)\right] \quad (2)$$

is the transverse field amplitude of the Gaussian embedded vortex beam of order, $l$[15]. Here, $m = 2\pi(n-1)s/\lambda$, is the phase contrast of the grating [19] with $n$ and $s$ being the refractive index and thickness of the lenslets of the MLA, respectively, at an input wavelength, $\lambda$, while $w_g$ being the beam waist radius of the Gaussian confined vortex beam. According to Fourier transformation theory [16], any object can be Fourier transformed by placing the object after the lens at an arbitrary distance, $d$, from the focal plane [20]. The basic principle of the technique is pictorially represented in Fig. 1, where the object (MLA) placed at a distance $(f - d)$ after the lens of focal length $f$, is Fourier transformed into array of vortices at the back focal plane for input vortex beams. Therefore, the field amplitude distribution of the vortex arrays at the back focal plane of lens, L3, can be obtained by taking the Fourier transform of Eq. (1) as [16]

$$E \sim \sum_{p,q=-\infty}^{\infty} J_{p,q}(m) \; \delta_{x,y}(d,\lambda) \; \otimes \; \mathcal{F}(E_l) \quad (3)$$

where,

$$\delta_{x,y}(d,\lambda) = \delta\left(\frac{1}{f\lambda}\left[x - p\left(\frac{\lambda d}{\Lambda}\right)\right], \frac{1}{f\lambda}\left[y - q\left(\frac{\lambda d}{\Lambda}\right)\right]\right), \quad (4)$$

$$\mathcal{F}(E_l) = \left[\frac{(r\sqrt{2})^l}{(w_0)^{l+1}} exp\left(\frac{-r^2}{w_0^2}\right) exp(il\varphi)\right], \quad (5)$$

$$J_{p,q}(m) = J_p\left(\frac{m}{2}\right) J_q\left(\frac{m}{2}\right). \quad (6)$$

As evident from Eq. 3, the convolution of $\mathcal{F}(E_l)$, with the 2D delta function, $\delta_{x,y}(d,\lambda)$, as represented by Eq. (4), replicates $\mathcal{F}(E_l)$ at positions determined by the individual delta peaks as a function of the input wavelength, $\lambda$, and distance, $d$. Therefore, one can generate array beams of desired intensity distribution by using suitable spatial profile of the input beam. Given that $\mathcal{F}(E_l)$, the Fourier transform of input Gaussian embedded vortex beam of order $l$, as presented in Eq. (5) resulting into a vortex beam with Gaussian beam waist, $w_0 = f\lambda/\pi w_g$, we expect an array of vortex beams at the Fourier plane of lens, L3. On the other hand, $J_{p,q}(m)$, the Bessel functions of first kind of orders $p$ and $q$ as represented in Eq. (6), determines the overall intensity distribution of the vortex array as a function of the phase contrast, $m$, of the grating. Therefore, by changing the value of $m = 2\pi(n-1)s/\lambda$, with the change of any of the parameters, $s$ and $\lambda$, one can control the overall intensity distribution of the vortex array. While the position of the individual vortex in the array is represented by the coordinates $(p,q)$ with $(0,0)$ being the central vortex, their relative intensities are determined by the value of $J_{p,q}(m)^2$. On the other hand, the pitch of the array given by $\Lambda_{exp} = \lambda d/\Lambda$, as shown in Eq. (4), is proportional to the distance of the MLA from the back focal plane of Fourier transform lens. Therefore, by simply adjusting, $d$, one can control the pitch of the vortex array. To confirm the control in overall intensity pattern of the vortex array, we have used a MLA of lenslet thickness, $s = 1.31$ μm and pitch, $\Lambda = 300$ μm and numerically calculated the variation in the intensity pattern of vortex arrays for input wavelengths, $\lambda = 1064$ nm and $\lambda = 532$ nm, corresponding to the phase contrasts, $m = 1.11\pi$ and $m = 2.27\pi$ respectively. The results are shown in Fig. 2. As evident from the first column, (a)-(b), of Fig. 2, we clearly see the change in the overall intensity pattern of the vortex array with the change in the phase contrast, $m$, of the MLA. Similarly, we have experimentally recorded the intensity pattern of the vortex array by using laser beams at 1064 nm and 532 nm as shown in second column, (c)-(d), of Fig. 2, in close agreement with the theoretical results. It is also interesting to note that, the decrease in the pitch, $\Lambda_{exp} = \lambda d/\Lambda$, of the vortex array with the decrease of the laser wavelength from 1064 nm to 532 nm, increases the total number of vortices in the array. In addition, the same intensity pattern can also be achieved by varying thickness, $s$, of the lenslets of MLA. Therefore, one can achieve vortex array beam of desired intensity pattern at a fixed wavelength by simply fabricating the lenslets of suitable thickness.

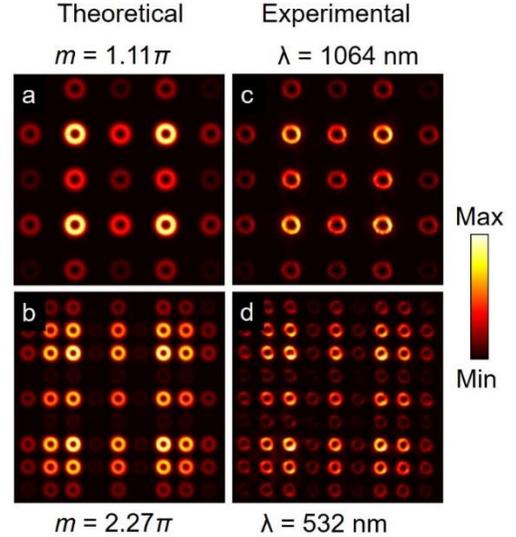

Fig. 2. (a)-(b) Theoretical and corresponding (c)-(d) experimental intensity distributions of vortex arrays of order $l = 3$ for different values of phase contrast, $m$. Experimentally the phase contrast, $m$ is varied using lasers of wavelength 1064 nm and 532 nm.

To verify the control in the pitch of the vortex array, we have Fourier transformed the input vortex beam of order $l = 1$ using the lens, L3, of focal length, $f = 300$ mm by keeping the MLA after the lens. Keeping the CCD camera fixed at the Fourier plane, we have recorded the intensity distribution of the vortex array for different positions, $d$, of the MLA away from the camera. As evident from Fig. 3(a)-(c), showing the intensity distribution of the vortex array for $d$ = 120 mm, 160 mm and 240 mm respectively, the separation of the individual vortices is increasing with the increase of $d$ without any change in the diameter of the vortices. To get further insight, we have measured the pitch and the diameter of the vortices in the array for different values of $d$ with the results shown in Fig. 3(d). As evident from Fig. 3(d), the pitch of the vortex array, $\Lambda_{exp}$, increases from 434 μm to 866 μm with the increase of MLA separation, $d$, from 120 mm to 240 mm. However, the diameter of the vortices remains constant around ~ 430 μm for all the positions of the MLA. While the upper limit of the pitch is decided by the focal length of the Fourier transforming lens, L3, the lower limit is decided by the mechnical constraint to position the MLA close to the Fourier plane and also the diameter of the vortices. Fitting the experimental results with the theoretical expression of the vortex pitch, $\Lambda_{exp} = \lambda d/\Lambda$, (solid line), we have experimentally measured the pitch of the MLA to be, $\Lambda$ = 298 μm close to the standard value, $\Lambda$ = 300 μm as provided by the manufacturer. Such observation clearly confirms the current technique as a simple and strightforward way of measuring the pitch of the MLAs.

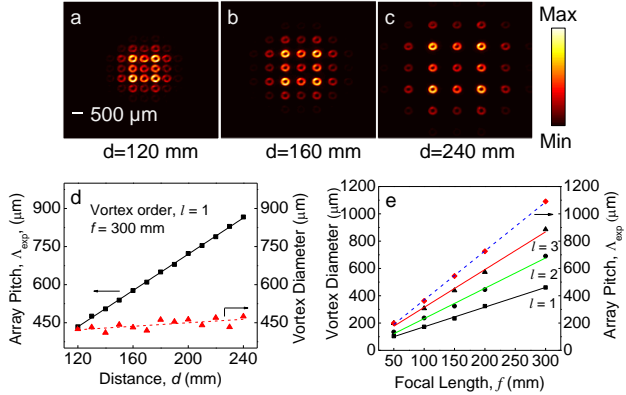

Fig. 3. (a)-(c) Intensity distribution of the vortex array of order $l$ = 1 for different position of the MLA. (d) Dependence of vortex beam diameter and pitch of the array on the distance '$d$' of MLA from the focal plane, (e) Variation of vortex beam diameter and pitch for input vortex orders, $l$ = 1, 2 and 3 on the focal length of the Fourier transforming lens. Lines are linear fit to the experimental results.

We have also verified the control in the size of the vortices of orders, $l$ = 1, 2 and 3 in the array by using a set of Fourier transforming lens, L3, of focal lengths, $f$ = 50, 100, 150, 200 and 300 mm. To avoid the mechanical constraint of keeping the MLA after the Fourier transforming lens especially for small focal lengths, we kept the MLA at a distance, $f$, before the lens and measured the size of the vortices at the Fourier plane. The results are shown in Fig. 3(e). As evident from Fig. 3(e), the diameter of the vortices of all orders, $l$ = 1, 2 and 3, vary from 103 μm, 134 μm, and 198 μm to 460 μm, 690 μm, and 886 μm, linearly with the variation of focal length of the Fourier transforming lens from, $f$ = 50 mm to $f$ = 300 mm. As expected, for a fixed focal length of the Fourier transforming lens, the size of the vortices in the array increases with the order, $l$, of the input vortices. Similarly, the pitch of the vortex array of all orders as represented by, $\Lambda_{exp} = \lambda f/\Lambda$, varies linearly with the focal length of the Fourier transforming lens. As shown in Fig. 3(e) (dashed line), the pitch varies linearly from 199 μm to 1091 μm with the focal length varing from $f$ = 50 to $f$ = 300 mm, independent to the order of the input vortex beam. Also, the experimentally measured value of MLA pitch from the best fit is observed to be, $\Lambda$ = 297 μm close to the standard value, $\Lambda$ = 300 μm, provided by the manufacturer, thus confirming the potential of the current technique. It is also interesting to note that, for a fixed focal length, the pitch of the array, $\Lambda_{exp} = \lambda d/\Lambda$, is independent of the focal length. Hence one can achieve wide range of pitch values by using a single lens of longer focal length.

With successful generation and control of the vortex arrays of different orders, we have also studied their frequency-doubling characteristics. Using the pump vortex array generated from the input vortex orders, $l_p$ = 2 and 6, at 1064 nm with constant power of P = 3.5 W, we have frequency-doubled the arrays into green at 532 nm by placing the BiBO crystal at the Fourier plane. To avoid the mechanical constraint, the exit plane is imaged by using the lenses, L4 and L5 in a 4$f$ ($f$ = 150 mm) imaging configuration and recorded using CCD. The results are shown in Fig. 4. Here we have shown nine spots of the arrays for better clarity without compromising on the scientific content of the experimental results. As evident from the first column, (a, b), of Fig. 4, both the arrays have beams with doughnut shaped intensity distribution. Using the tilted lens technique[21], splitting the vortices into their characteristic lobes, $n_p = l_p + 1$, as shown in the second column (c, d), of Fig. 4, we confirm that the vortices of each array have same order and sign. The order of the vortex arrays is measured to be, $l_p$ = 2 and 6, same as that of the input vortex. Similarly, the corresponding second harmonic (SH) beam have intensity pattern same as that of the pump vortex array with nine doughnut shaped beams as shown in third column, (e, f) of Fig. 4. The tilted lens technique confirms the order of the SH vortex array as shown in fourth column, (g, h), of Fig. 4, to be $l_{sh}$ = 2 x $l_p$ = 4 and 12 respectively, twice the order of the pump vortices satisfying the OAM conservation in the SHG process.

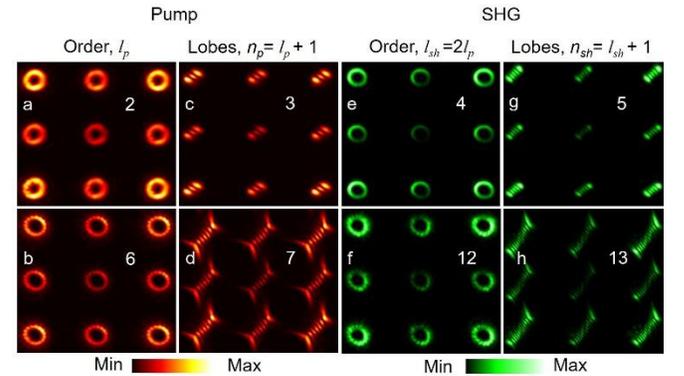

Fig. 4. Intensity distribution of (a, b) vortex beams, and corresponding (c, d) lobe structure generated through the tilted lens of the pump vortex arrays of orders, $l_p$ = 2 and $l_p$ = 6. Corresponding (e, f) intensity distribution and (g, h) lobe structure of the SH vortex array.

We have also measured the focusing dependent SHG efficiencies of vortex arrays of different orders. Keeping the pump power fixed at P = 3.5 W, we have pumped the crystal with vortex array of orders, $l_p$ = 0(Gaussian beam), 1, 2, and 3, using a set of lenses of focal lengths, $f$ = 25, 50, 100, 150, 200 and 300 mm and recorded the SHG power. The results are shown in Fig. 5(a). As evident from Fig. 5(a), like optical array of Gaussian beams ($l_p$ = 0), the vortex array of all orders follow similar focusing dependent SHG efficiency showing maximum SHG efficiency for the focusing lens of $f$ = 50 mm. Since the pitch of the vortex array does not change with its order, the decrease in SHG efficiency with the order of the vortex array at a fixed focusing condition can be attributed to the increase in the size of the vortex beam in the array (see Fig. 3(e)). Using the lens, $f = 50$ mm, the maximum single-pass SHG efficiency of the vortex array of orders, $l_p = 0$(Gaussian beam), 1, 2, and 3, are measured to be 11.5 %, 3.6 %, 1.8 %, and 1.1 %, respectively. We have also measured the power scalability of the SH vortex array source. Focusing the vortex array of order, $l_p =1$, using the lens of, $f = 50$ mm, we measured the single-pass SHG power and efficiency as a function of pump power with the results shown in Fig. 5(b).

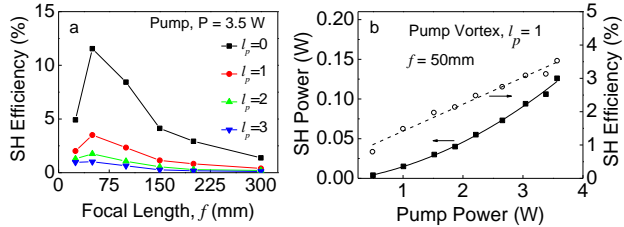

Fig. 5. (a) Variation of SH efficiency of vortex arrays of different orders with the focal length of the focusing lens. Lines are guide to the eyes. (b) Dependence of SH vortex array power and efficiency on the power of the pump vortex array of order, $l_p = 1$. Solid and dotted linear are quadratic and linear fit to the experimental results.

As evident from Fig. 5(b), the SHG power and efficiency of the vortex array show a quadratic and linear dependence, respectively, to the input pump vortex array power. The maximum SHG vortex array power is measured to be 138 mW at 3.8 W of pump power corresponding to a single-pass conversion efficiency as high as ~3.65%. However, no saturation effect observed in the conversion efficiencies, indicating the possibility of increased SHG power with further increase in the pump power.

In conclusion, we have demonstrated a generic experimental scheme based on a MLA and a Fourier transforming lens to generate high power, ultrafast, array beams with spatial intensity pattern same as the input beam. As a proof-of-principle, using input vortex beam of order as high as $l = 6$, we have generated vortex beam arrays. By placing the MLA after the Fourier transforming lens and subsequently adjusting the position of the MLA and focal length of the Fourier lens, we have controlled the size and pitch of the vortex beams in the array. Further, using single-pass SHG of the vortex array at 1064 nm we have generated vortex array at 532 nm with vortex order as high as 12. We also observe the vortex array beam to follow the focusing dependent SHG efficiency similar to the Gaussian beam. The maximum power of the SH vortex array beam is measured to be 138 mW of order, $l_p= 1$ at a single-pass SHG efficiency of ~ 3.65%.